\documentclass[12pt]{iopart}

\usepackage{graphicx}
\usepackage{verbatim}
\usepackage{bm}

\newcommand{\beq}{\begin{equation}}
\newcommand{\eeq}{\end{equation}}
\newcommand{\M}{\mathcal{M}}

\begin{document}

\title[Delay Regulated Explosive Synchronization in Multiplex Networks]{Delay Regulated Explosive Synchronization in Multiplex Networks}

\author{Ajay Deep Kachhvah$^1$ \& Sarika Jalan$^{1,2}$}

\address{1. Complex Systems Lab, Indian Institute of Technology Indore - Simrol, Indore - 453552, India}
\address{2. Centre for Biosciences and Biomedical Engineering, Indian Institute of Technology Indore - Simrol, Indore - 453552, India}
\ead{sarikajalan9@gmail.com}
\vspace{10pt}

\begin{abstract}
It is known that explosive synchronization (ES) in an isolated network of Kuramoto oscillators with inertia is significantly enhanced by the presence of time delay. Here we show that time delay in one layer of the multiplex network governs the transition to synchronization and ES in the other layers. 
We found that a single layer with time-delayed intra-layer coupling, depending on the values of time delay, experiences a different type of transition to synchronization, e.g., ES or continuous, and the same type of transition is incorporated simultaneously in other layer(s) as well. Hence, a suitable choice of time-delay in only one layer can lead to a desired (either ES or continuous) transition simultaneously in the delayed and other undelayed layers of multiplexed network. These results offer a platform for a better understanding of the dynamics of those complex systems which are represented by multilayered framework and contain time delays in the communication processes.
\end{abstract}

%
\vspace{2pc}
%
\submitto{\NJP}
%
%
%

\section{Introduction}\label{intro}
\noindent The networks have been proven to be an important tool to investigate manifold processes such as synchronization, epidemic spreading, transport and communication in complex systems \cite{barrat,camy}. In recent years, a novel synchronization process exhibiting hypersensitivity or explosiveness in the natural physical and biological systems, called explosive synchronization (ES), has drawn much attention in the scientific community \cite{gomez,leyva,leyva3,leyva4,bocca,peron,peron2,arenas}. The epileptic seizures in the brain \cite{adhikari}, chronic pain in the Fibromyalgia brain \cite{lee}, the cascading failure of power grids \cite{buldy}, and the jamming of the Internet \cite{huberman} have shown ES transitions with small initial perturbations. Further, it is known that time-delay is naturally present in communication systems and enhances synchronization in the corresponding systems \cite{yeung,choi,kandel,dhamala,ares,herrgen}. The effect of time delay on ES transition has been demonstrated in the Kuramoto model having the degree-frequency correlation, and in the second-order Kuramoto model \cite{peron,ajay}.
  
  Lately, extensive studies of ES transition on single layer networks have been carried out. However, in many complex systems several dynamics take place at the same time and to understand these systems in better manner an isolated network structure turns out to be insufficient and hence, a multilayered structural approach is required. In the multilayered approach, same set of the nodes, representing a complex system's units, are laid out in several discrete layers, each one of them representing a different type of interaction process existing among the nodes. In the multilayered networks, the synchronization of different dynamical processes evolving simultaneously in individual layer may get affected by structural and dynamical properties of other layers leading to different characteristics for the synchronization experienced by the layer. Relay, cluster and delay induced synchronization have been reported in multiplex architectures \cite{leyva2,asingh,asingh2}. Of late, the emergence of ES has been shown in the multilayer networks comprised of Kuramoto oscillators as well as second-order Kuramoto oscillators \cite{zhang,nicosia,ajay2}.
  
  \begin{figure}[t]
	\begin{center}
		\includegraphics[height=6cm,width=16cm]{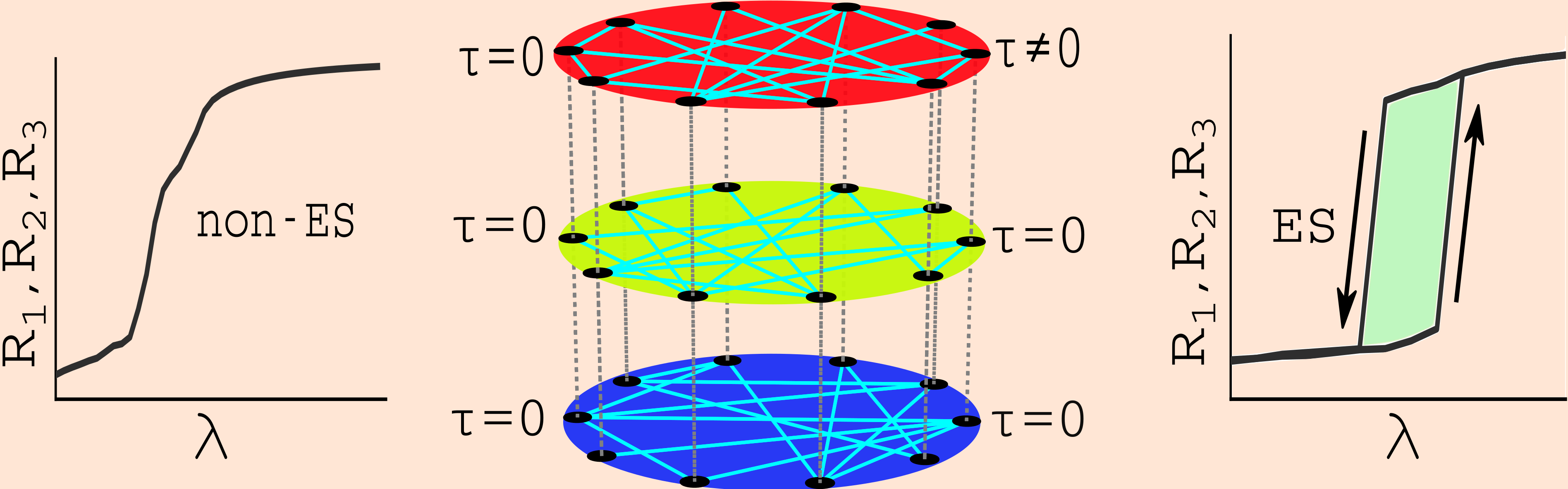}\\
	\caption{(Color online) Schematic diagram of a 3-layered multiplex network with intra-layer and inter-layer connections depicted by solid lines (cyan) and dashed lines (grey), respectively. It illustrates the effect of absence and presence of time delay in only one layer on nature of synchronization transition in all the layers in multiplex network. For un-delayed case, all the three layers manifest a regular second-order transition, however, as a time delay is employed in a single layer (layer $1$), all the three layers experience ES transition simultaneously for certain value of delay. $\lambda$ depicts over-all coupling strength, and $R_1,R_2\ \mbox{and}\ R_3$ provide means of global synchronization as defined in \sref{model}.}
		\label{figure1}
	\end{center}
\end{figure}
  In this paper, we discuss about the repercussions of inclusion of time delay on explosive synchronization in the multilayer networks. We represent the nodes in multiplex networks by second-order Kuramoto oscillators \cite{tanaka,fila,dorf,pinto,trees} with natural frequencies positively correlated with their respective degrees. The time delay is introduced in one layer and its impact on synchronization in other layers are studied. It is perceived that inclusion of time delay in only one layer induces either ES or non-ES transition concurrently in all the layers depending on the value of the time-delay. It is further revealed that the occurrence of ES or non-ES transition is solely governed by the evolution of average frequency of the system. Such occurrences of ES or non-ES transition were first revealed for the single layer time delayed networks \cite{peron,ajay}. The important revelation coming out from our study is that by making an appropriate choice of time delay in one layer, a desired type of transition can be achieved concurrently in all the multiplexed layers. The impact of various structural and dynamical properties on the observed transition behavior has been discussed in details both numerically and analytically.

\section{Delayed Multiplex Network Model}\label{model}

\noindent In the current study, we investigate how the presence of time-delay between the pair of nodes in a single layer affects transitions to synchronization in other layers of multiplex networks. Here, we consider undirected multiplex networks where nodes in each layer are represented by the second order Kuramoto oscillators with one layer subject to time delayed coupling while rest to regular coupling between each pair of nodes (\fref{figure1}). We examine the impact of time-delayed layer on transition to synchronization in other un-delayed layers.
The multiplex network we consider is comprised of $L$ layers of networks having same number of the nodes $N$.  
\begin{figure}[t]
	\begin{center}
		\includegraphics[height=8cm,width=16cm]{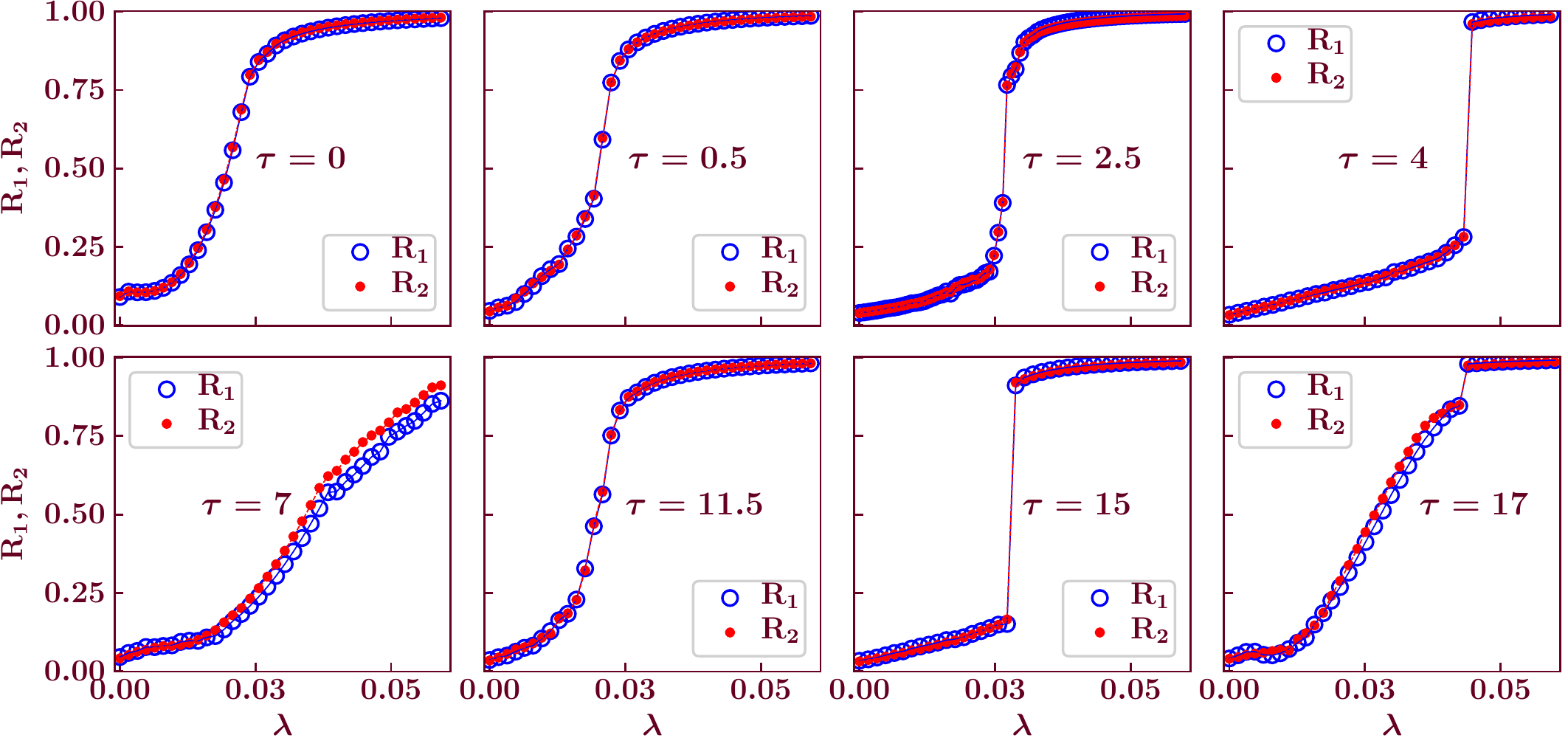}\\
	\caption{(Color online) Synchronization transitions in multiplex networks comprised of two ER networks having $\langle k_1\rangle=\langle k_2\rangle=12$, for different values of the time-delay introduced in layer $1$.}
		\label{figure2}
	\end{center}
\end{figure}
The time-evolution dynamics of the nodes, denoted by second order Kuramoto oscillators, in the presence or absence of the time-delay in a layer are governed by
\begin{eqnarray} \label{eq:evol}
\fl m\ddot\theta^i_{\alpha}(t) {+} \dot\theta^i_{\alpha}(t) = \Omega^i_{\alpha}(0) {+} \lambda_{\alpha} \sum_{j{=}1}^{N} A^{i,j}_\alpha \sin[\theta^j_\alpha(t-\tau)-\theta^i_{\alpha}(t)]+\sum_{\beta=1}^{L-1}\sigma_{\alpha\beta}\sin[\theta^i_{\beta}(t)-\theta^i_{\alpha}(t)], \nonumber \\
\fl m\ddot\theta^i_{\beta}(t) {+} \dot\theta^i_{\beta}(t) = \Omega^i_\beta(0) {+} \lambda_{\beta} \sum_{{j=}1}^{N} A^{i,j}_\beta \sin[\theta^j_\beta(t)-\theta^i_{\beta}(t)]+\sum_{\beta=1}^{L-1}\sigma_{\beta\alpha}\sin[\theta^i_{\alpha}(t)-\theta^i_{\beta}(t)],
\end{eqnarray}
respectively, where $i=1,...,N$, $\Omega_{i}(0)$ ($\theta_{i}(t)$) is the natural frequency (the instantaneous phase) of $i$th oscillator, parameter $m$ is mass, $\lambda$ is coupling strength and $\tau$ is time-delay introduced between the phases of nodes in a layer. Superscript $\alpha$ denotes the single delayed layer while $\beta\neq\alpha$ denotes all the other $L-1$ un-delayed layers. $\sigma_{\alpha\beta}=\sigma_{\beta\alpha}$ denotes inter-layer coupling strength between layer $\alpha$ and $\beta$. The multiplex network is represented by $L$ layers encoded by a set of adjacency matrices, $\{\bi{A}_1,..., \bi{A}_\kappa,..., \bi{A}_L \}$, where
element $\bi{A}_\kappa^{ij}=1$ when the nodes $i$ and $j$ are connected and $\bi{A}_\kappa^{ij}=0$, otherwise. Adjacency matrix (size $NL\times NL$) of the multiplex network can be written in block matrix form as following
\begin{equation} \label{eq:adj}
 \M=\left(
\begin{array}{cccc}
\bi{A}_1    & \sigma_{12}\bi{I} & \cdots & \bi{O} \\ 
\sigma_{21}\bi{I} & \bi{A}_2    & \cdots & \bi{O} \\
\vdots & \vdots & \ddots & \vdots \\ 
\bi{O} & \bi{O} & \cdots & \bi{A}_L
\end{array}
\right)
\end{equation}
where $\bi{I}$ and $\bi{O}$ are identity and null matrices, respectively.

We quantify transition to synchronization in a layer by means of the order parameter. We define order parameters $R_\alpha$ and $R_\beta$ at time $t$ for the time-delayed and regularly coupled layer as
\begin{eqnarray} \label{eq:op}
 R_\alpha(t)e^{\iota\psi_\alpha(t-\tau)}=\frac{1}{N}\sum_{j=1}^{N}e^{\iota\theta^j_\alpha(t-\tau)},\nonumber \\
 R_\beta(t)e^{\iota\psi_\beta(t)}=\frac{1}{N}\sum_{j=1}^{N}e^{\iota\theta^j_\beta(t)},
\end{eqnarray}
where $\psi(t)$ denotes average phase of the nodes. Hence, $R$ is a measure of coherence of the collective dynamics of the nodes, i.e., the degree of synchronization of the network. $R=1$ corresponds to a completely synchronized state, while $R=0$ denotes a asynchronous state. 
\begin{figure}[t]
	\begin{center}
		\includegraphics[height=6cm,width=12cm]{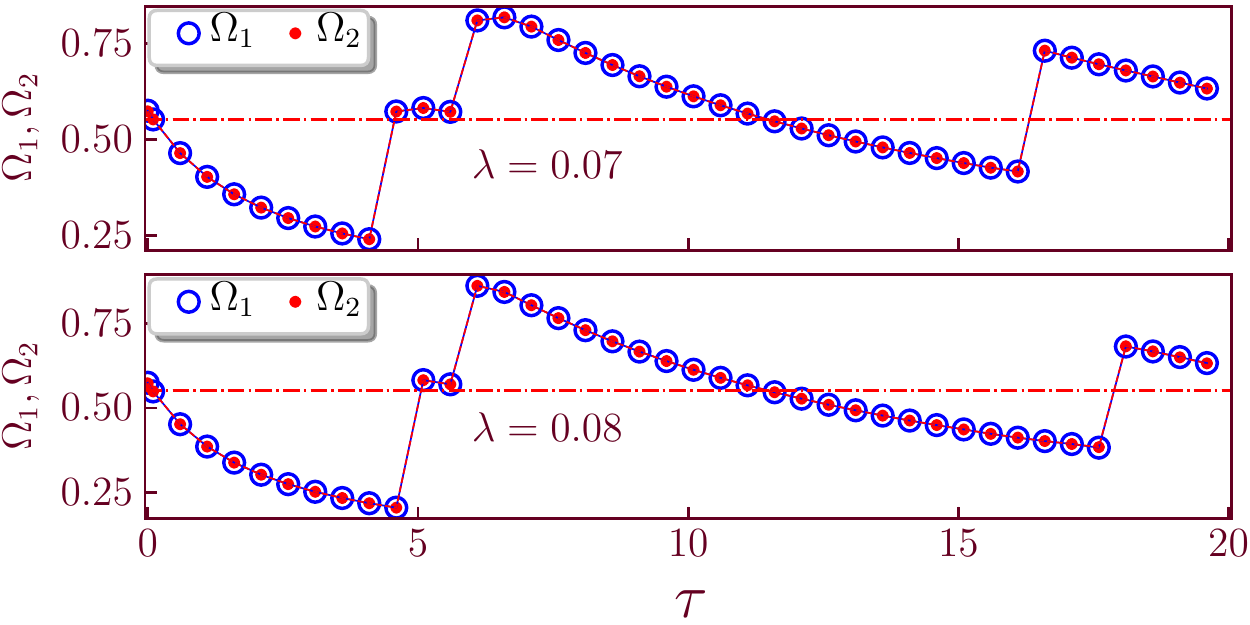}\\
		\caption{(Color online) Average frequencies $\Omega_{1}$ and $\Omega_{2}$ for the two ER layers of the multiplex network as a function of time-delay $\tau$ corresponding to different coupling strengths.}
		\label{figure3}
	\end{center}
\end{figure}

\section{Simulation results} \label{num}
Here, we present and discuss in details various simulations results about the effect of time delay in a single layer on synchronous transitions in the other layers of multiplex networks. First, we compute the order parameters defined in \eref{eq:op} as a function of the coupling strength to track the transition characteristics of the network layers. Here note that, while computing various dynamical quantities such as order parameters, frequencies, etc., each layer of the multiplex network is subject to the same coupling strength $\lambda=\lambda_\alpha=\lambda_\beta=\lambda_\gamma,\cdots$.
We adiabatically increase the coupling strength $\lambda$ from $\lambda_0$ (incoherent) to $\lambda_0{+}n\delta\lambda_0$ (synchronous) state in the step of $\delta\lambda_0$.
We also adiabatically decrease the coupling strength from $\lambda_0{+}n\delta\lambda_0$ (synchronous) to $\lambda_0$ (incoherent) in the step of $\delta\lambda_0$, and compute the order parameter $R$ for each layer at each value of $\lambda$. Before each $\delta\lambda_0$ step, we eliminate initial transients and integrate the system long enough ($10^4$ time steps) using a fourth-order Runge-Kutta method with time step $dt=0.01$, to arrive at the stationary state. For all the simulations, initial phases for the nodes in the individual layer are drawn from a random uniform distribution in the range $[0, 2\pi)$. We have made a specific choice for the natural frequencies, the natural frequencies of the nodes in each layer are set equal to their respective degrees, i.e., $\Omega^i(0)=k^i/k^{max}$, where $k^i=\sum_{j=1}^N A^{ij}$ is the degree of a node and $k^{max}$ is maximum degree in the network. All the simulation results presented in this paper are for $N=100$ nodes in each layer constructed with average degree $\langle k\rangle=12$ for various network topologies, otherwise mentioned elsewhere.
 
\begin{figure}[t]
 	\begin{center}
 		\hspace{-0.5cm}
 		\includegraphics[height=10cm,width=16cm]{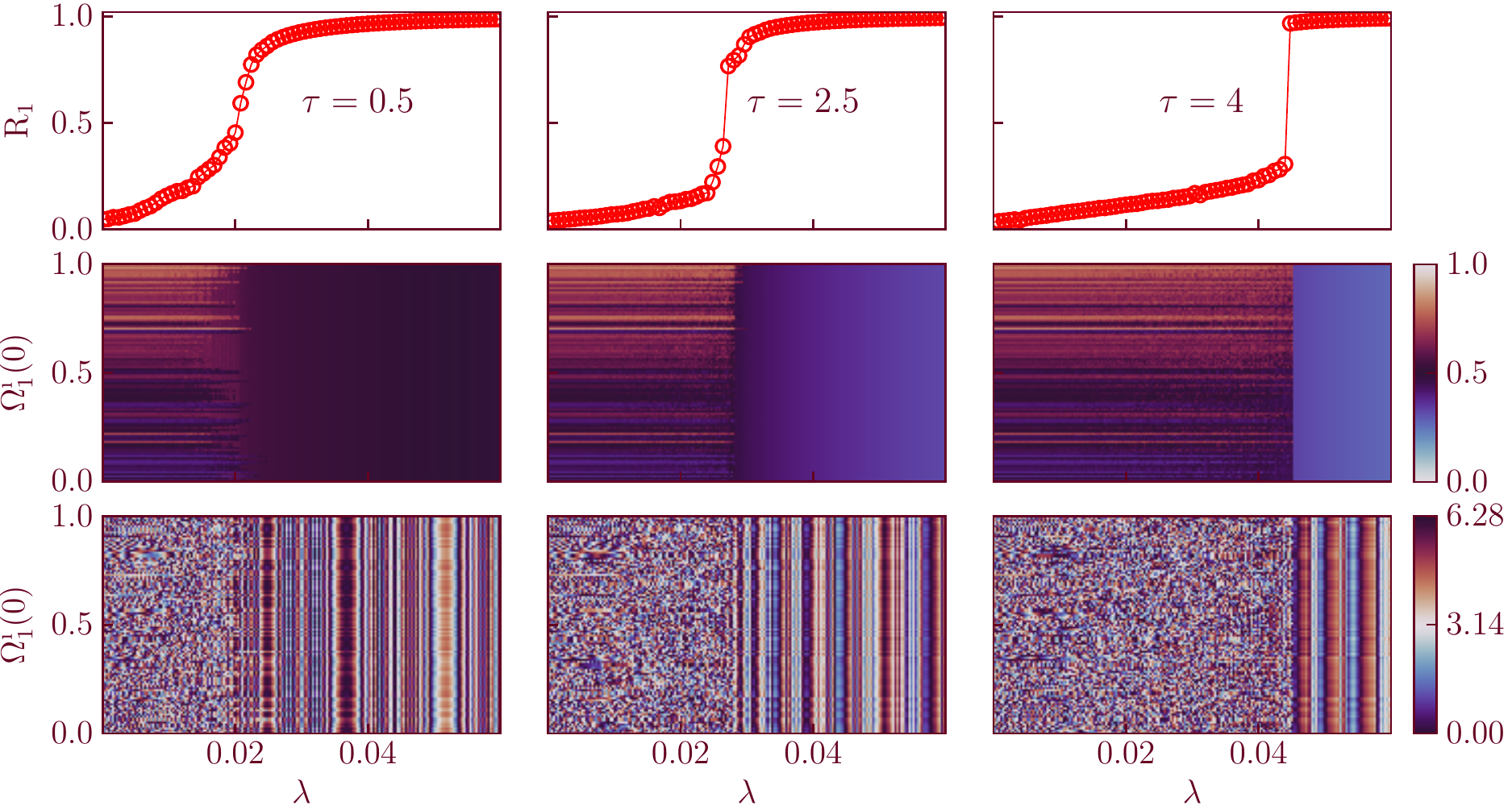}\\
 		\caption{(Color online) Order parameter $R_1$, effective frequency $\Omega^i_{1}$ and steady-state phases $\theta^i_{1}$ of nodes of the first layer of the multiplex network comprised of two ER networks. Left, middle and right panels correspond to the time-delay $\tau=0.5, 2.5\ \mbox{and}\ 4$, respectively.}
 		\label{figure4}
 	\end{center}
\end{figure} 
For numerical study, firstly, we consider a duplex network comprised of two ER random networks \cite{erdos} of average degrees $\langle k_1\rangle=\langle k_2\rangle=12$ and introduce a constant time delay between the phases of nodes in layer $1$. We compute the order parameters for both the layers in \fref{figure2}.
In the absence of time-delay, i.e., $\tau=0$, both the ER layers display second-order transitions and synchronize simultaneously. However, the presence of time delay significantly enhances the synchronization behavior of the multiplexed layers. For different values of the time delay, different transition behaviors, i.e., either strong or weak ES or second-order is witnessed in both the layers. For an instance in \fref{figure2}, rather a weak ES is seen for $\tau=2.5$ and $17$, strong ES for $\tau=4$ and $\tau=15$, and very weak ES or second-order transition for $\tau=0.5$, $\tau=7$ and $\tau=11.5$, respectively. Moreover, both the layers synchronize simultaneously exhibiting the same nature of transition. The onset of transition to synchronization for different values of $\tau$ is witnessed at different critical coupling strength. 
From these observations one can infer that introduction of a time-delay not only enhances transition in the layer that includes the time delay, but also the same nature of transition is incorporated in the other layer(s) simultaneously via inter-layer coupling. Hence in the multiplex networks, a desired (ES or continuous) transition can be incorporated simultaneously in all the layers by making a suitable choice of the time-delay.

\begin{figure}[t]
	\begin{center}
		\hspace{-0.5cm}
		\includegraphics[height=8cm,width=16cm]{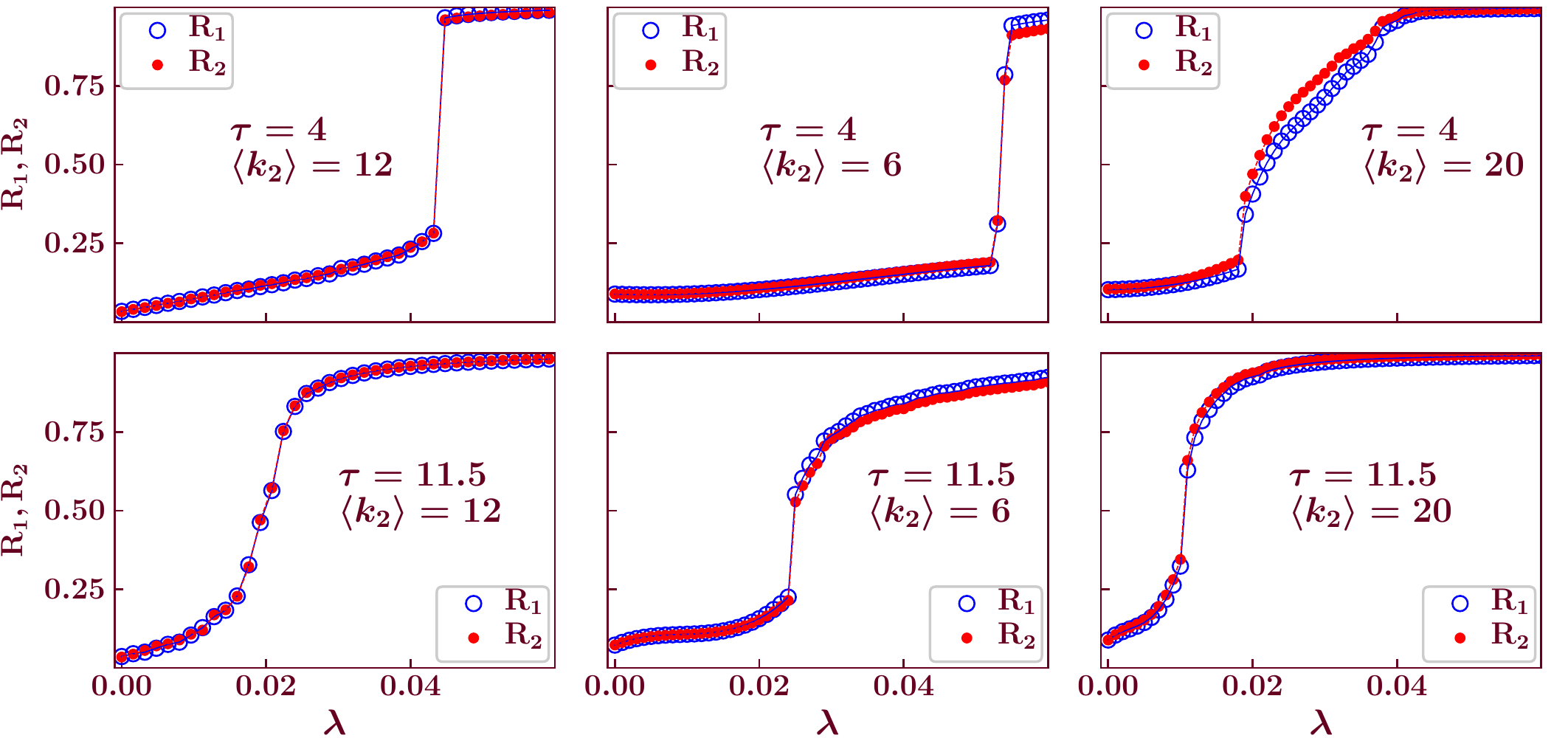}\\
		\caption{(Color online) Synchronization transitions when average degree of the first layer is fixed at $\langle k_1\rangle=12$ and that of the second layer is varied to a lower and higher values.}
		\label{figure5}
	\end{center}
\end{figure}
A dependency of average frequency $\Omega$ on the time delay $\tau$, i.e., $\Omega=\Omega(\tau)$ is responsible for such interesting behavior.
Therefore, to understand the underlying microscopic dynamics behind such time-delay dependent behavior observed in a multiplex network, we define effective frequency $\langle{\Omega^i}\rangle$ of each node and average frequency $\Omega$ of the network as 
\begin{eqnarray}
\langle{\Omega^i}\rangle=\frac{1}{T}\int_t^{t+T}\dot\theta_i(t)dt,\\
\Omega=\frac{1}{N}\sum_{i=1}^N \langle{\Omega^i}\rangle,
\end{eqnarray}
where $T$ is the total time of the averaging after eliminating the initial transients. We compute average frequencies $\Omega_{1}$ and $\Omega_{2}$ for the two layers as a function of time-delay $\tau$ (see \fref{figure3}). The initial (at t=0) average frequencies of the first and the second layers are $\Omega^0_{1}=0.575$ and $\Omega^0_{2}=0.573$, respectively.
In the presence of time-delay, the average frequencies of both the layers $\Omega_{1}$ and $\Omega_{2}$, simultaneously exhibit an oscillatory behavior around $\Omega^0_{1}$ and $\Omega^0_{2}$ (\fref{figure3}). Starting from $0$, as the value of $\tau$ is increased, $\Omega_{1}$ and $\Omega_{2}$ exhibit a gradual decrease. Interestingly, at a certain value of $\tau$, $\Omega_{1}$ and $\Omega_{2}$ suddenly jump to higher values and again start decreasing along a new branch as $\tau$ is increased further.
\begin{figure}[t]
	\begin{center}
		\hspace{-0.5cm}
		\includegraphics[height=4cm,width=12cm]{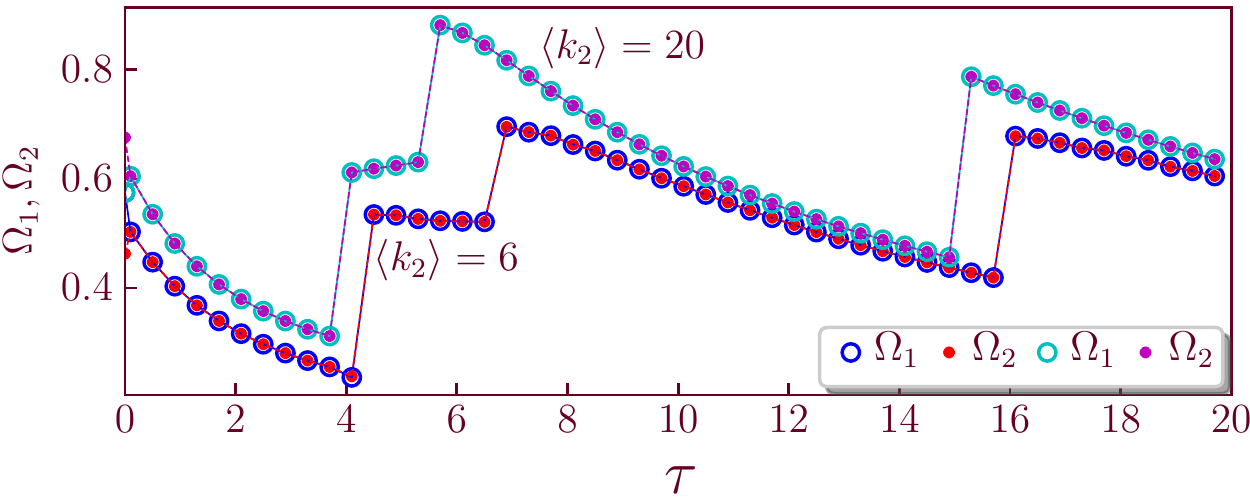}\\
		\caption{(Color online) Average frequencies $\Omega_1$ and $\Omega_2$ of the two multiplexed ER layers as a function of the time-delay $\tau$. Here results correspond to coupling strength $\lambda=0.06$, $\langle k_1\rangle=12$ and $\langle k_2\rangle$ being lower and higher than $\langle k_1\rangle$.}
		\label{figure6}
	\end{center}
\end{figure}

From \fref{figure3}, it is apparent that the value of $\tau$ for which differences $|\Omega_{1}-\Omega^0_{1}|$ and $|\Omega_{2}-\Omega^0_{2}|$ are large, a strong ES is observed in both the layers (see \fref{figure2} for $\tau=4$ and $\tau=15$). While for both the differences being at intermediate values, a rather weak ES is observed (see \fref{figure2} for $\tau=2.5$ and $\tau=17$). Moreover, if both the differences are low, a second-order transition is observed in both the layers (see \fref{figure2} for $\tau=0.5$ and $\tau=11.5$). Hence, such oscillatory dependence of $\Omega_{1}$ and $\Omega_{2}$ on the time delay $\tau$ causes different type of transitions in the multiplexed layers.

\begin{figure}[t]
	\begin{center}
		\hspace{-0.5cm}
		\includegraphics[height=8cm,width=16cm]{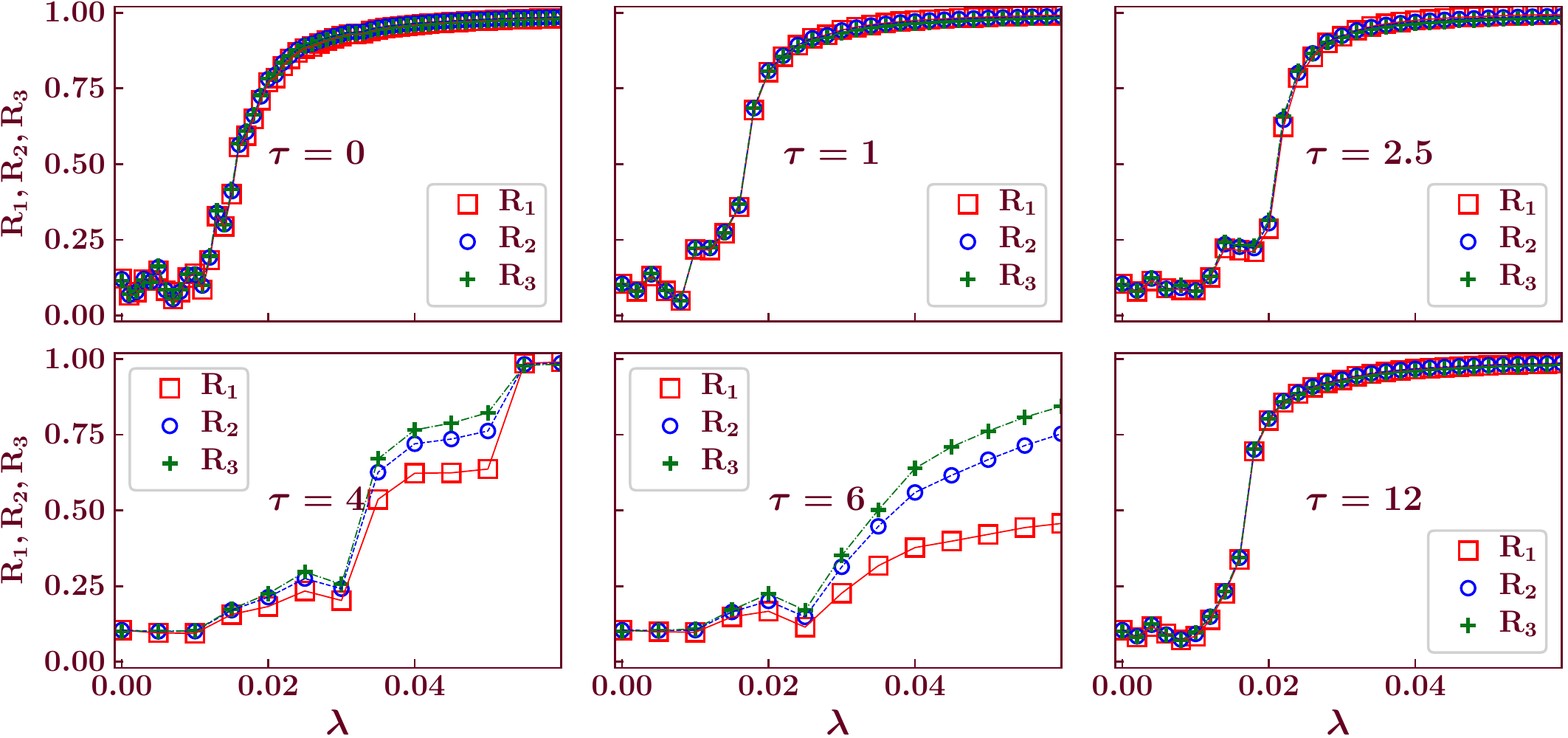}\\
		\caption{(Color online) Multiplex network comprised of three ER networks of N=$100$ nodes each and $\langle k_1\rangle=\langle k_2\rangle=\langle k_3\rangle=12$. The order parameters $R_1, R_2\ \mbox{and}\ R_3$ for different time delays employed only in the first layer.}
		\label{figure7}
	\end{center}
\end{figure}
To further understand the microscopic dynamics of ES with time delay, we plot the order parameter $R_1$ along with the corresponding effective frequencies of nodes $\langle{\Omega^i_1}\rangle$ and steady-state phases $\theta^i_{1}$ associated with the layer $1$ in \fref{figure4} as both the layers synchronize simultaneously displaying the same transitional behavior for a certain value of the time delay. It is apparent that for $\tau=4$, ES transition takes place as effective frequency $\langle{\Omega^i_1}\rangle$ and, in turn, phases $\theta^i_{1}$ get locked to their respective average values together at the critical coupling strength (see \fref{figure4}). For $\tau=2.5$, most of the nodes are locked to their respective average frequencies and average phases at the critical coupling strength, however, rest of the nodes get locked later at a bit higher coupling strength. For $\tau=0.5$ (see \fref{figure4}), $\langle{\Omega^i_1}\rangle$ and $\theta^i_{1}$ gradually start reaching to their respective stationary values near the critical coupling strength, thereby leading to the continuous transitions.
\begin{figure}[t]
	\begin{center}
		\hspace{-0.5cm}
		\includegraphics[height=8cm,width=16cm]{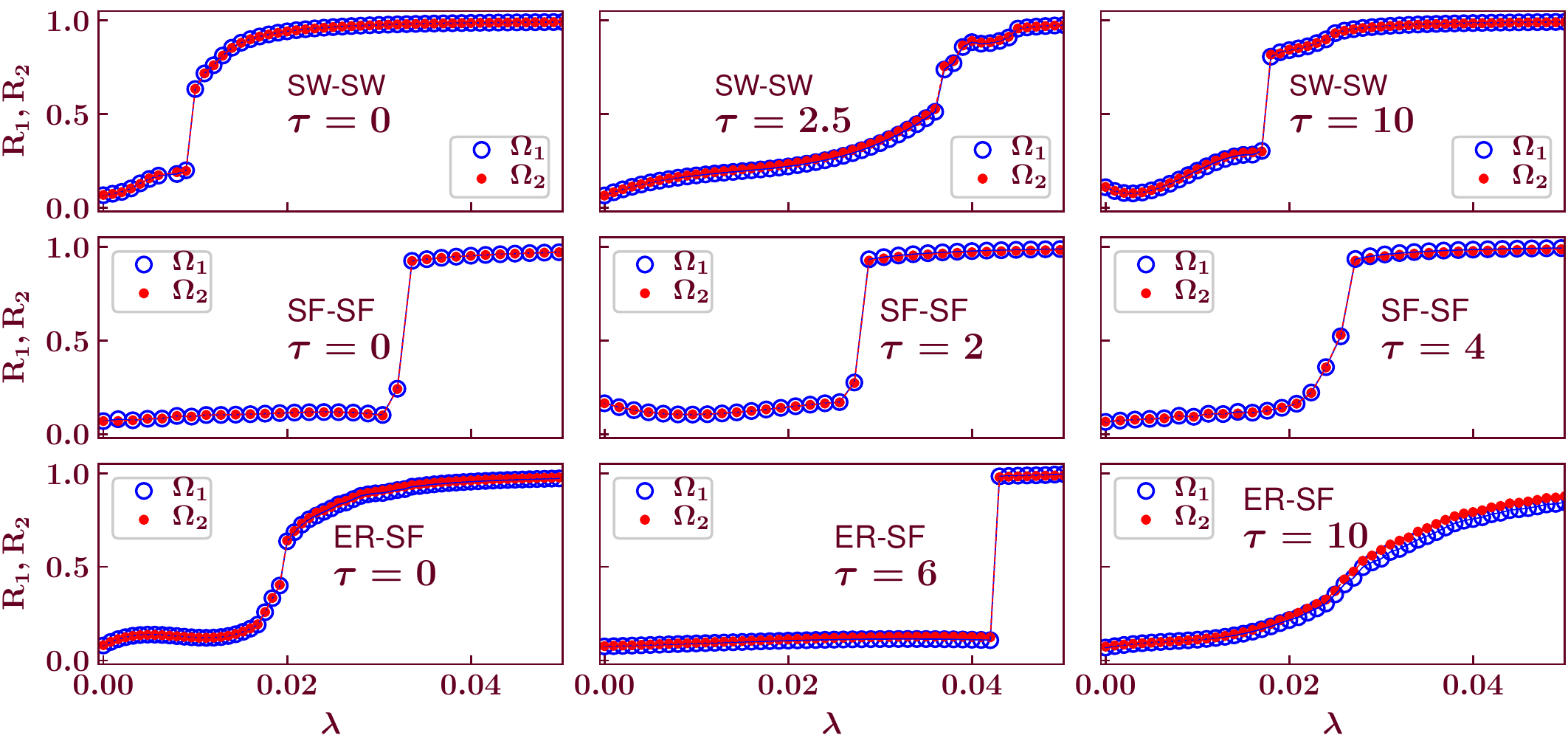}\\
		\caption{(Color online) Multiplex networks comprised of two SW, two SF and ER-SF networks with $\langle k_1\rangle=\langle k_2\rangle=12$. Synchronization transitions for different time-delay $\tau$ employed in the first layer for the each network-pair of the multiplex networks (ER layer in case of ER-SF network).}
		\label{figure8}
	\end{center}
\end{figure}

\subsection{Effect of average degree of the multiplexed layers on nature of transition:} Next, we study the effect of different average degree of the network layers on synchronization transition. To investigate this, we fixed the average degree of the ER layer $1$ to a value $\langle k_1\rangle$ and chose an equal, a lower and a higher value for the average degree $\langle k_2\rangle$ of the ER layer $2$ (\fref{figure5}). For $\tau=4$, ES and second-order transitions corresponding to $\langle k_2\rangle=6\ \mbox{and}\ 20$, respectively, are observed at different critical coupling strengths as compared to that of $\langle k_2\rangle=12$. In the same fashion, weak ES transitions for $\tau=11.5$ corresponding to $\langle k_2\rangle=6\ \mbox{and}\ 20$ are observed at different critical coupling strengths as compared to second-order transition observed for $\langle k_2\rangle=12$.
Therefore, the variation in the average degree $\langle k_2\rangle$ of the non-delayed layer shifts the critical coupling strength which, in turn, enhances or subsides the transition accordingly. Different oscillatory dependence of $\Omega_1$ and $\Omega_2$ on $\tau$, corresponding to $\langle k_2\rangle=6\ \mbox{and}\ 20$ (see \fref{figure6}), accounts for this behavior.

\subsection{Effect of number of multiplexed layers $L$ on the transition:}
It is also important to verify whether the same set of observations hold for the multiplex networks having more than two layers. In order to investigate this, we consider a multiplex network consisting of three ER layers and compute the order parameter for each layer for different values of the time delay. From \fref{figure7}, it is quite apparent that for each value of the time delay $\tau$, a different path to the synchronization, i.e., ES or continuous transition takes place. It is to be noted that even though $\mathcal M$ follows the connection matrix \eref{eq:adj}, which indicates that the delayed layer $1$ and the un-delayed layer $3$ are not directly interconnected, yet, by means of multiplexing via layer $2$, the layer $3$ exhibits exactly the same transition type as that of the layer $1$. Hence, this is an instance of distant coordination similar to the relay synchronization in the multiplex networks \cite{leyva2}. On the basis of this finding we can say that for a multiplex network of a finite number of layers, the intra-layer delay introduced in one layer is sufficient to incorporate the same transition behavior in rest of the layers. However, the oscillatory dependence of average frequency on delay for all the layers changes as the number of layers $L$ increases, which, in turn, changes the critical coupling strength and corresponding transition type for all the layers concurrently. For an instance, transitions corresponding to $\tau=2.5$ and $\tau=4$ are observed at different coupling strengths for the 2-layered (see \fref{figure2}) and the 3-layered (see \fref{figure7}) multiplex networks.

\subsection{Synchronization on a variety of multiplex networks:}
\begin{figure}[t]
	\begin{center}
		\includegraphics[height=6cm,width=14cm]{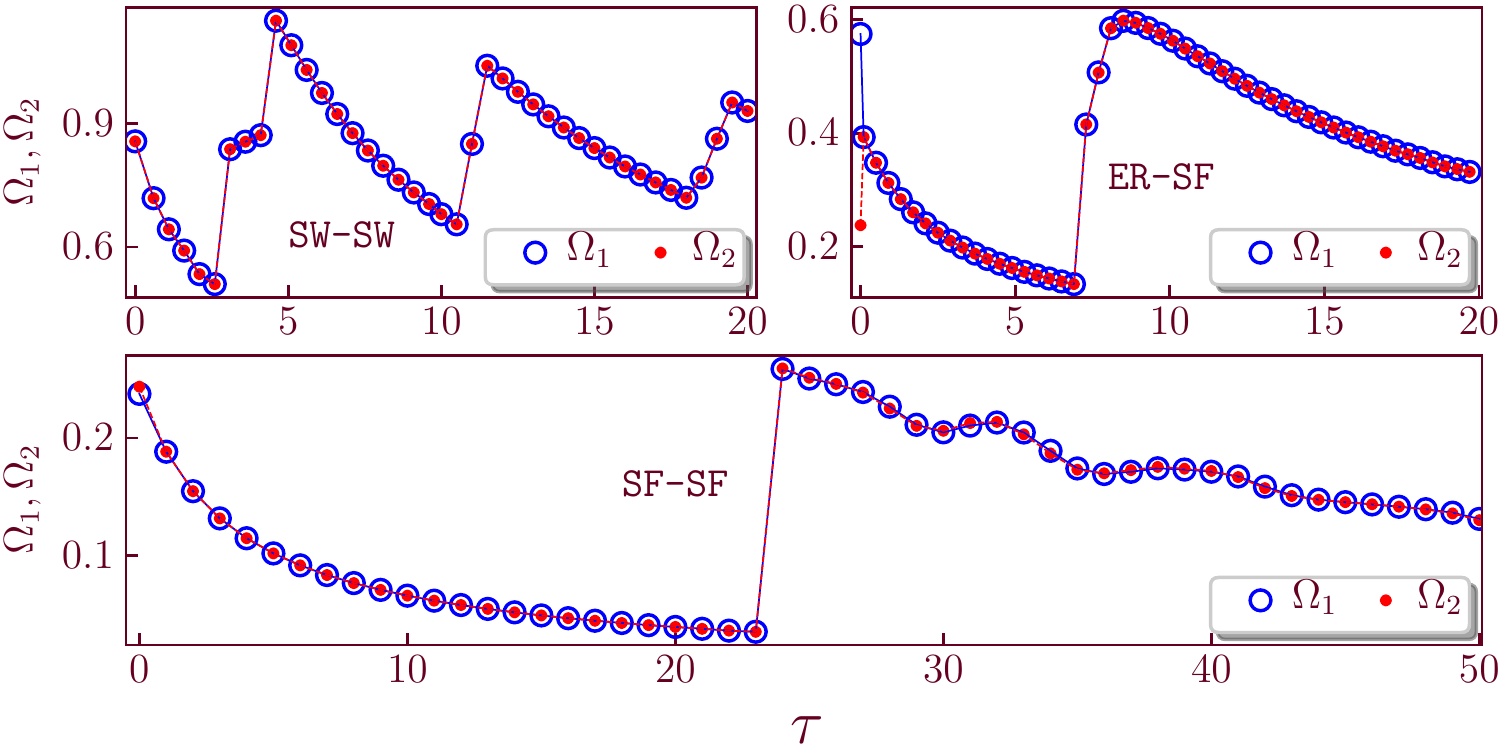}\\
		\caption{(Color online) Average frequencies $\Omega_{1}$ and $\Omega_{2}$ of the two multiplexed SW-SW, ER-SF and SF-SF layers as a function of time-delay $\tau$ corresponding to the coupling strength $\lambda=0.06$.}
		\label{figure9}
	\end{center}
\end{figure}

Next, we investigate the transition behavior in the presence of time delay on a variety of multiplex networks constructed from selecting different network topologies. To explore this, we consider multiplex networks comprised of two small-world (SW) \cite{watts}, two scale-free (SF) \cite{barabasi}, and an ER and a SF networks having time delay introduced in one layer (in ER layer in case of ER-SF multiplexed layers). We follow the same analysis as performed for the case of two ER layers. In the absence of time delay ($\tau=0$) (see \fref{figure8}), the two SW layers, the two SF layers and the ER-SF layers exhibit ES, strong ES, and weak ES transitions, respectively, in contrary to the second-order transition observed in the case of two ER layers. However, depending on the value of time delay either a strong ES, weak ES or second-order transition is observed for each selected network-pair of the multiplexed layers (see \fref{figure8}). To fathom the role of average frequency of each layer in determining the transition behavior of the selected pairs of multiplexed layers, we compute the average frequencies $\Omega_1$ and $\Omega_2$ for the layers as a function of the time delay as shown in \fref{figure9}. 
For these cases too, an oscillatory dependence of average frequencies $\Omega_{1}$ and $\Omega_{2}$ for the two layers on time delay $\tau$ is witnessed. In the case of two SF multiplexed layers, the observed oscillatory branches of average frequencies $\Omega_1$ and $\Omega_2$ are stretched further along the axis $\tau$ as compared to those belonging to the other pairs of the multiplexed layers.

Therefore, it is manifested that the time delay plays an important role in determining the nature of transition in a variety of multiplexed networks. We have shown that in multiplexed networks, time delay incorporates same intra-layer transition properties in all of the layers. In the presence of time delay, difference in the average connectivity of the multiplexed layers also seems to either strengthen or subside ES transition in the layers.

\section{Analytical treatment}
In \sref{num}, we have witnessed that the departure of average frequency $\Omega$ from its initial value $\Omega_0$ determines the nature of transition to synchronization. Additionally, it is numerically demonstrated that the average frequency $\Omega$ of a multiplexed layer depends not only on the time delay $\tau$ but also on the average degree $\langle k\rangle$ of all the layers. In this section we make an attempt to elucidate analytically the dependence of average frequency $\Omega$ on dynamical as well as structural properties of the multiplex networks. Therefore, to obtain average frequency, we rewrite \eref{eq:evol} in the terms of the order parameters $R_\alpha(t)$ and $R_\beta(t)$ using \eref{eq:op} and by defining degree of a node as $k^i=\sum_{j=1}^NA^{ij}$
\begin{eqnarray} \label{eq:evol2}
\fl	m\ddot\theta^i_{\alpha}(t) {+} \dot\theta^i_{\alpha}(t) = \Omega^i_{\alpha}(0) {+} N\lambda_{\alpha}k_{\alpha}^iR_{\alpha} \sin[\psi_{\alpha}(t-\tau)-\theta_{\alpha}^i(t)] + \sum_{\beta=1}^{L-1}\sigma_{\alpha\beta}\sin[\theta^i_{\beta}(t)-\theta^i_{\alpha}(t)], \nonumber\\
\fl	m\ddot\theta^i_{\beta}(t) {+} \dot\theta^i_{\beta}(t) = \Omega^i_{\beta}(0) {+} N\lambda_{\beta}k_{\beta}^iR_{\beta} \sin[\psi_{\beta}(t)-\theta_{\beta}^i(t)] + \sum_{\beta=1}^{L-1}\sigma_{\alpha\beta}\sin[\theta^i_{\alpha}(t)-\theta^i_{\beta}(t)].   
\end{eqnarray}
In the stationary state, $R_{\alpha}(t)\simeq R_{\alpha}$ and $R_{\beta}(t)\simeq R_{\beta}$. Next, we consider a separate rotating frame for each layer with average phase $\psi_\alpha(t)=\Omega_\alpha t$. The new variables in the rotating frame are then defined as $\phi_\alpha^i=\theta_\alpha^i-\psi_\alpha$. Sets of equations \eref{eq:evol2} get transformed to the following in the rotating frames
\begin{eqnarray} \label{eq:evol3}
\fl	m\ddot\phi_{\alpha}^i+\dot\phi_{\alpha}^i=[\Omega_{\alpha}^i(0)-\Omega_{\alpha}] - N\lambda_{\alpha}R_{\alpha}k_{\alpha}^i\sin[\phi_{\alpha}^i+\Omega_{\alpha}\tau] + \sum_{\beta=1}^{L-1}\sigma_{\alpha\beta}\sin[\phi_{\beta}^i-\phi_{\alpha}^i+(\Omega_{\beta}-\Omega_{\alpha})t],\nonumber \\
\fl	m\ddot\phi_{\beta}^i+\dot\phi_{\beta}^i=[\Omega_{\beta}^i(0)-\Omega_{\beta}] - N\lambda_{\beta}R_{\beta}k_{\beta}^i\sin[\phi_{\beta}^i] + \sum_{\beta=1}^{L-1}\sigma_{\alpha\beta}\sin[\phi_{\alpha}^i-\phi_{\beta}^i+(\Omega_{\alpha}-\Omega_{\beta})t].
\end{eqnarray}

In synchronous state, it is safe to assume that all the nodes in each layer are locked to their respective mean field $\psi_{\alpha}$, i.e., $\psi_{\alpha}=\theta_{\alpha}^i$. Therefore, summing over the set of $N$ equations for each layer, \eref{eq:evol3} yields the following relations for the average frequencies 
\begin{eqnarray} \label{eq:mf}
\Omega_{\alpha} = \Omega^0_{\alpha} - N\lambda_{\alpha}R_{\alpha}\langle k_{\alpha}\rangle\sin[\Omega_{\alpha}\tau]+ \sum_{\beta=1}^{L-1}\sigma_{\alpha\beta}\sin[(\Omega_{\beta}-\Omega_{\alpha})t],\nonumber\\
\Omega_{\beta} = \Omega^0_{\beta} - \sum_{\beta=1}^{L-1}\sigma_{\alpha\beta}\sin[(\Omega_{\beta}-\Omega_{\alpha})t],
\end{eqnarray}
where $\Omega^0_{\alpha}=\frac{1}{N}\sum_{i=1}^N \Omega^i_{\alpha}(0)$ and $\Omega^0_{\beta}=\frac{1}{N}\sum_{i=1}^N \Omega^i_\beta(0)$ are initial $(t{=}0)$ average frequencies, and average degree $\langle k_\alpha\rangle=\frac{1}{N}\sum_{i=1}^Nk^i_{\alpha}$. For the sake of convenience, we present our analysis to a two layered multiplex network, i.e., $L=2$. For a duplex network, \eref{eq:mf} can be expressed as
\begin{eqnarray} \label{eq:mf12}
\Omega_{1} = \Omega^0_{1} - N\lambda_{1}R_{1}\langle k_{1}\rangle\sin[\Omega_{1}\tau]+ \sigma_{12}\sin[(\Omega_{2}-\Omega_{1})t],\nonumber\\
\Omega_{2} = \Omega^0_{2} - \sigma_{12}\sin[(\Omega_{2}-\Omega_{1})t].
\end{eqnarray}
Further, we define $\Delta\Omega$, i.e., the difference between $\Omega_{1}$ and $\Omega_{2}$ as 
\begin{equation} \label{eq:fq}
\Delta\Omega = \Omega_{2}-\Omega_{1} = \Delta\Omega_0 + \Omega_{1}^d(\tau,\lambda_{1}) - 2\sigma_{12}\sin[\Delta\Omega.t],
\end{equation}
where constant $\Delta\Omega_0=\Omega^0_{2}-\Omega^0_{1}=\langle k_2\rangle/k_2^{max} - \langle k_1\rangle/k_1^{max}$, and $\Omega_1(\tau,\lambda_1)=N\lambda_{1}R_{1}\langle k_1\rangle\sin[\Omega_{1}\tau]$ is also a constant as $\Omega_1(\tau)$ and $R_1$ would be constant in the stationary state for a certain $\tau$ and $\lambda_1>\lambda_1^c$. In the absence of time delay, i.e., $\tau=0$, \eref{eq:fq} reduces to 
\begin{equation} \label{eq:fq0}
\Delta\Omega = \Delta\Omega_0 - 2\sigma_{12}\sin[\Delta\Omega.t].
\end{equation}
If $\Omega^0_1$ and $\Omega^0_2$ are equal, that, in turn, implies $\Delta\Omega_0\simeq0$, consequently, \eref{eq:fq0} becomes 
\begin{equation}
\Delta\Omega = -2\sigma_{12} \sin[\Delta\Omega.t]
\end{equation}
It is quite apparent from \eref{eq:mf12} or \eref{eq:fq} that inter-dependent $\Omega_1$ and $\Omega_2$ depend on parameters $\tau$, $\lambda$, $\sigma_{12}$, $\langle k_1\rangle$ and $\langle k_2\rangle$, which is in further concurrence with the numerical results shown in \sref{num} that the average degree of the layers along with the time delay determine the characteristic of transition in the multilayered framework.

\section{Discussions}
In the present work, we have discussed the influence of time delay in a single layer on the nature of phase transition in all the layers of multiplex networks.
We demonstrate that path by which multiplexed layers achieve synchronous states, can be controlled by the means of the time delays. The multiplexed layers altogether choose a common path, either a strong or weak ES or a second-order one, to reach to the respective synchronous states, which is solely determined by the value of time delay introduced in a single layer. Each value of the time delay departs the average frequencies of all the layers from their initial values leading to different but nearby values, and the extent of such departure governs the type of transition which is common to all the layers. If the extent of the departure is large, this leads to ES or first-order transition in all the layers. An intermediate departure gives rise to weak ES, whereas a small departure only results in a second-order transition. Our this finding holds true for a variety of multiplex networks constructed from different network topologies. However, in the case of SF-SF multiplex network, no second-order transition is observed, only ES or weak ES transition exists. This might be due to the strong heterogeneity of the SF layers, which favors the ES transition in presence of degree-frequency correlation. In presence of time delays, the difference in average connectivities of the layers brings upon changes in the behavior of oscillatory dependency of the average frequencies on time-delay which, in turn, may result in a different type of transition. It is also demonstrated that for the 3-layered multiplex networks with no direct inter-layer connections between the remote layers, one delayed layer is capable of inducing the same type of transition in the remotely un-delayed layers too, demonstrating that multiplex networks favor remote synchronization. We have also shown analytically that the average frequency of each layer depends on the inter-layer coupling strength and the average connectivity of all the layers along with the time-delay.

 Lately, the evidence of ES is reported in the brains of patients suffering from fibromyalgia (FM), a condition described as an omnipresent, chronic pain \cite{lee}. The patients with FM have brain networks predisposed towards abrupt, global responses or abnormal hypersensitivity (ES) to minor changes. ES is worth investigating as it is detrimental to certain biological and physical circumstances, e.g., epileptic seizure in which  things are abruptly turned on \cite{adhikari} and the power grid failure in which things are abruptly turned off \cite{buldy}. 
Investigation of ES on real world complex systems by adopting multilayer (or multiplex) architecture has more potential as the multilayer approach provides a more accurate representation of many real world complex systems. 

Researchers have employed multiplex perspective to identify the most important functional region of the brain. A human functional brain network consists of peripheral and central or hub regions operating at different frequency bands, which can be modeled as multiplex network in which each layer is identified by a single frequency band carrying unique topological information \cite{domenico, buldu}. 
Information processing from various functional or sensory regions of the brain is a cooperative process of neurons and time delay is naturally incorporated in the information processing to form a coherent and melded perception of the outer real world. Hence, time-delayed synchronization (ES and non-ES) of neurons, while employing multiplex network perspective in the representation of various functional sensory regions, in information processing is worth investigating. 
Hence, our study on ES on the multiplex framework can be prolific in not only providing better understanding but also regulating the dynamics of interwoven concurrent dynamical processes occurring in both biological and physical systems, specifically the brain network.

\ack
SJ acknowledges DST, Government of India, project grant EMR/2016/001921, and BRNS - DAE, Government of India, project grant 37(3)/14/11/2018-BRNS/37131 for their financial support.

\end{document}